\documentclass[12pt]{article}
\usepackage{graphicx}
\usepackage{amssymb, amsmath}

\newcommand\pubnumber{SNSN-323-63}
\newcommand\pubdate{\today}
\usepackage{hyperref}
\def\institute{University of Hamburg\\
Luruper Chaussee 149, 22761 Hamburg, Germany}
\def\support{\footnote{E-mail: nataliia.kovalchuk@desy.de}}

\def\Title#1{\begin{center} {\Large #1 } \end{center}}
\def\Author#1{\begin{center}{ \sc #1} \end{center}}
\def\Address#1{\begin{center}{ \it #1} \end{center}}

\newcommand\pubblock{\rightline{\begin{tabular}{l} \pubnumber\\
         \pubdate  \end{tabular}}}
\newenvironment{Abstract}{\begin{quotation}  }{\end{quotation}}
\newenvironment{Presented}{\begin{quotation} \begin{center} 
             PRESENTED AT\end{center}\bigskip 
      \begin{center}\begin{large}}{\end{large}\end{center} \end{quotation}}

\def\beq{\begin{equation}}
\def\eeq#1{\label{#1}\end{equation}}
\def\eeqn{\end{equation}}

\def\beqa{\begin{eqnarray}}
\def\eeqa#1{\label{#1}\end{eqnarray}}
\def\eeqan{\end{eqnarray}}

\let\bar=\overbar

\def\Dslash{\not{\hbox{\kern-4pt $D$}}}
\def\dslash{\not{\hbox{\kern-2pt $\del$}}}

\def\msb{{\bar{\ssstyle M \kern -1pt S}}}

\begin{document}
\begin{titlepage}
\pubblock

\vfill
\Title{Measurement of the top quark mass with lepton+jets final states in pp collisions at $\sqrt{s}=13~\mathrm{TeV}$}
\vfill
\Author{Nataliia Kovalchuk on behalf of the CMS Collaboration\support}
\Address{\institute}
\vfill
\begin{Abstract}
The mass of the top quark is measured using a sample of $\rm{t}\bar{\rm t}$ candidate events with one lepton, muon or electron, and at least four jets in the final state, collected by the CMS detector in pp collisions at $\sqrt{s}=13~\mathrm{TeV}$ at the CERN LHC. The candidate events are selected from data corresponding to an integrated luminosity of $35.9~\mathrm{fb}^{-1}$. For each event the mass is reconstructed from a kinematic fit of the decay products to a $\rm{t}\bar{\rm t}$ hypothesis. The top quark mass is determined simultaneously with an overall jet energy scale factor (JSF), constrained by the mass of the W boson in $\rm{q}\bar{\rm q}$ decays. The measurement is calibrated on samples simulated at next-to-leading order matched to parton shower. The top quark mass is found to be $172.25\pm 0.08\,\rm{(stat+JSF)} \pm 0.62\,\rm{(syst)}~\mathrm{GeV}$. 
The dependence of this result on event kinematical properties is studied and compared to predictions of different models of $\rm{t}\bar{\rm t}$ production.
\end{Abstract}
\vfill
\begin{Presented}
$10^{th}$ International Workshop on Top Quark Physics\\
Braga, Portugal,  September 17--22, 2017
\end{Presented}
\vfill
\end{titlepage}
\def\thefootnote{\fnsymbol{footnote}}
\setcounter{footnote}{0}

\section{Introduction}
The first top quark mass ($m_{\rm{t}}$) measurement in lepton+jets final state using the full data sample collected in 2016 by the CMS detector\,\cite{Chatrchyan:2008aa} at a centre-of-mass energy $\sqrt{s}$ = 13\,TeV is presented~\cite{CMS-PAS-TOP-17-007}, corresponding to an integrated luminosity of $35.9 \pm 0.9~\mathrm{fb}^{-1}$~\cite{CMS-PAS-LUM-17-001}. The implementation and evaluation of the ideogram method~\cite{Abdallah:2008xh}, which is used for the $m_{\rm{t}}$ extraction utilizing the Monte Carlo (MC) simulated events, is performed, analogously to the mass extraction technique employed for the most precise CMS measurement~\cite{Khachatryan:2015hba} using Run\,1 data. 

Since the publication of the CMS Run\,1 measurements, new versions of Monte Carlo generators have been integrated and a larger data set has been collected at $\sqrt{s}=13$\,TeV. The new simulations allow more refined estimations of the modeling uncertainties, while the analysis techniques and final state are the same as in~\cite{Khachatryan:2015hba}. Additionally, the dependence of the measured mass value on the kinematic properties of the events is evaluated. 

\vspace{-5pt}
\section{Data Samples and Event Selection}

The measurement is performed by analyzing events in top quark pair production (t$\bar{\rm{t}}$) with lepton+jets in the final state, $\rm{t}\bar{\rm t} \rightarrow \rm Wb Wb \rightarrow (\rm{l}\rm{\nu_{\rm{l}}}\rm{b})(\rm{q}\bar{\rm q}\rm{b})$, which combines both $\mu$+jets and $e$+jets decay channels. Hence, events are required to pass a single-muon trigger with a minimum threshold on the transverse momentum ($p_{\rm{T}}$) of an isolated muon of $24$\,GeV or a single-electron trigger with a $p_{\rm{T}}$ threshold for isolated electrons of $32$\,GeV.

Simulated $\rm{t}\bar{\rm{t}}$ signal events are generated with $\textsc{powheg}$~v2~\cite{Nason:2004rx,Campbell:2014kua} and the $\textsc{pythia}$~8.219 parton-shower generator~\cite{Sjostrand:2007gs} using the \textsc{CUETP8M2T4} tune~\cite{Skands:2014pea} for seven different top quark mass values: 166.5, 169.5, 171.5, 172.5, 173.5, 175.5, and  178.5\,GeV. 

Events with exactly one isolated muon with $p_{\rm{T}}>26$\,GeV and $| \eta |<2.4$ or exactly one isolated electron with $p_{\rm{T}}>34$\,GeV and $| \eta |<2.1$ are selected. In addition, at least four jets with $p_{\rm{T}}>30$\,GeV and  $| \eta |<2.4$ are required. Additionally, exactly two jets originating from the b quarks are required, being clustered with the CSVv2~\cite{CMS:2016kkf} algorithm that combines reconstructed secondary vertices and track-based lifetime information~\cite{Chatrchyan:2012jua}. 

\begin{figure*}[!htb]
\center
\includegraphics[width=0.39\textwidth]{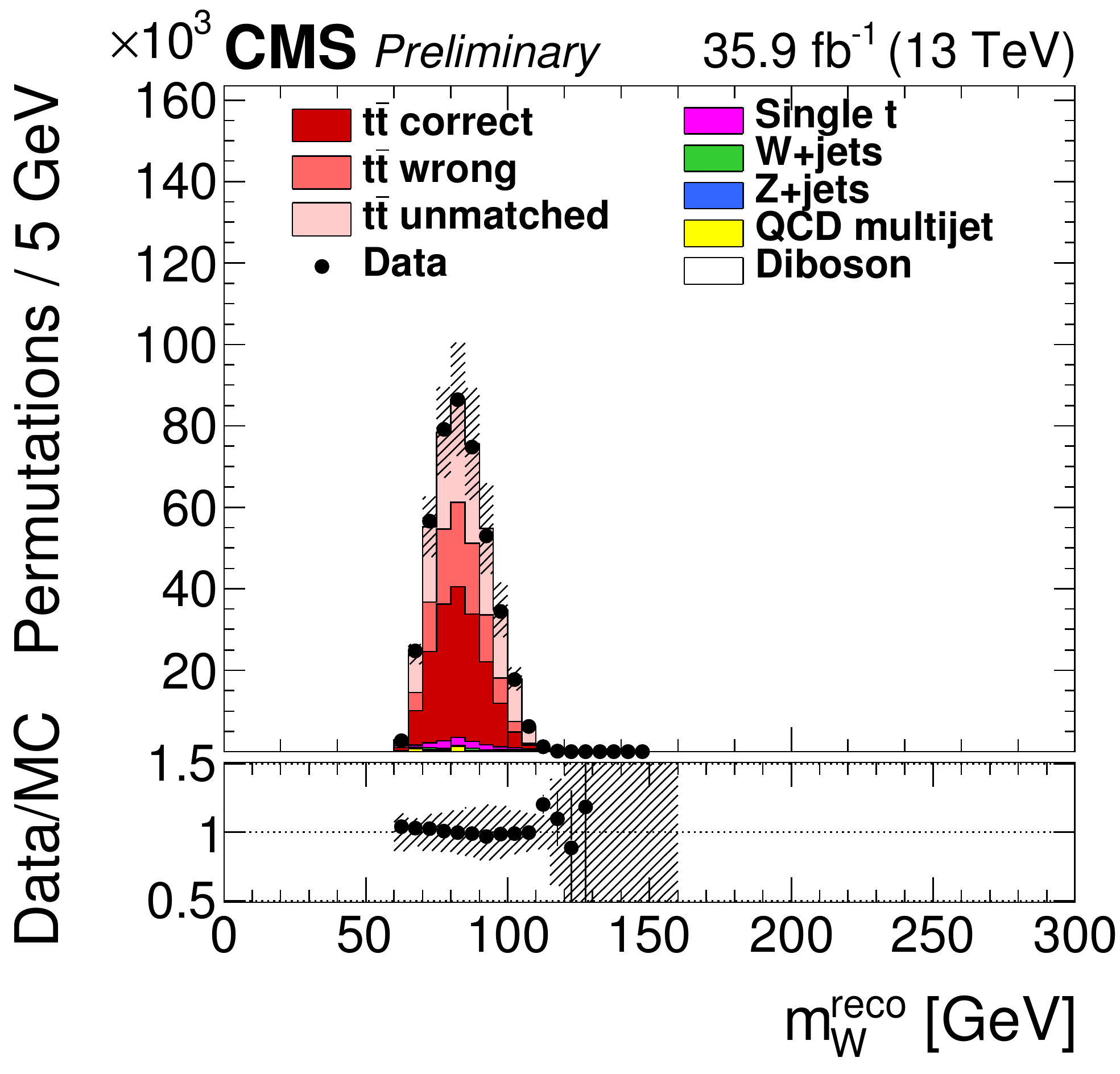}
\includegraphics[width=0.39\textwidth]{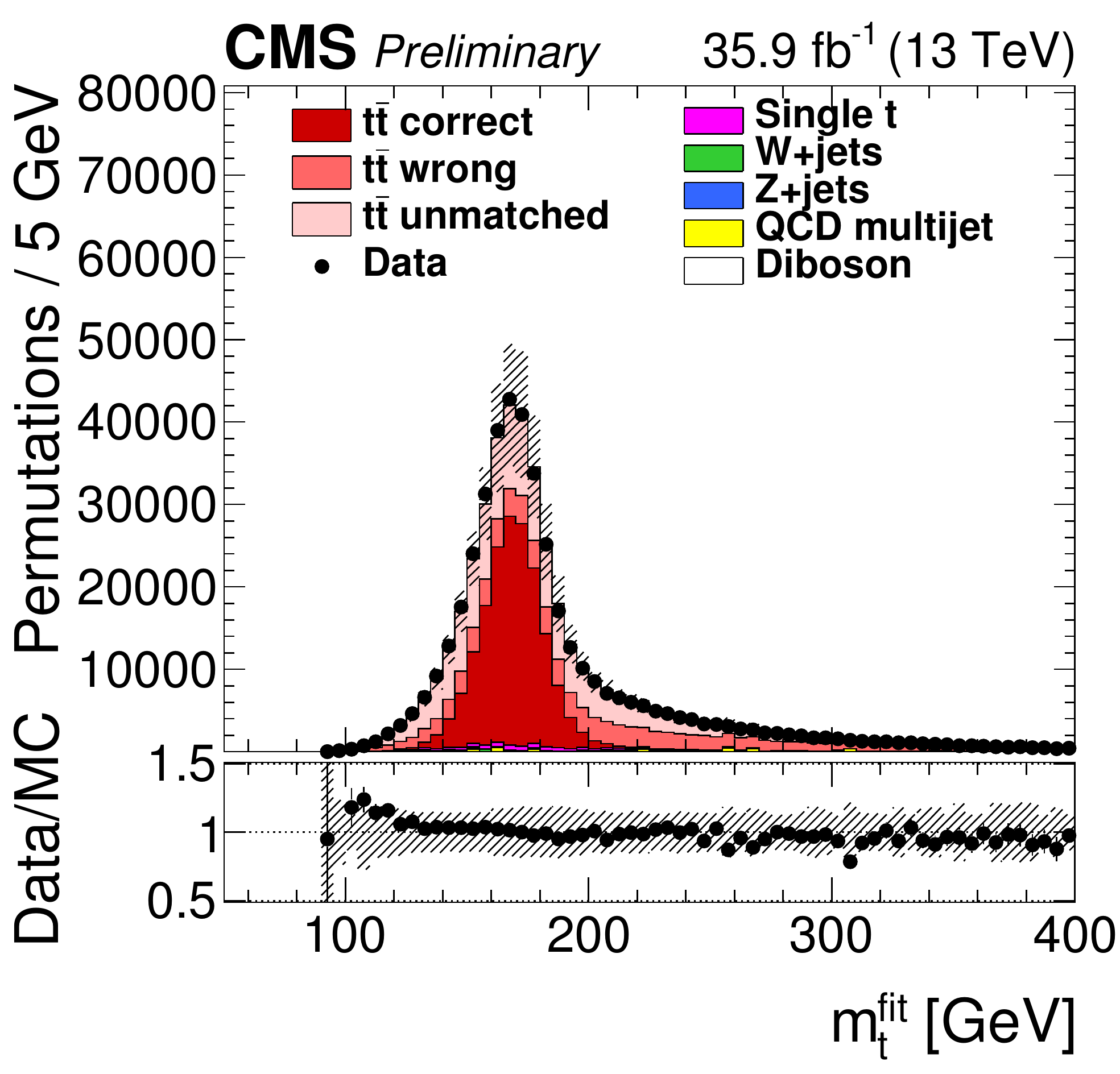}

\caption{\label{fig:controlplot-weighted-obs}
(left) The reconstructed W boson masses $m_{\rm{W}}^{\rm{reco}}$ and (right) the fitted top quark masses $m_{\rm{t}}^{\rm{fit}}$ after the goodness-of-fit selection and the weighting by $P_\mathrm{gof}$.
The vertical bars show the statistical uncertainty and the hatched bands show the statistical and systematic uncertainties added in quadrature. The lower portion of each panel shown the ratio of the yields between the collision data and the simulation. Figure taken from~\cite{CMS-PAS-TOP-17-007}.}
\end{figure*}

For the final selection, the events are reconstructed and the parton-jet assignments for simulated $\rm{t}\bar{\rm{t}}$ events are classified
as \emph{correct} ($\mathrm{cp}$), \emph{wrong} ($\mathrm{wp}$), and \emph{unmatched} ($\mathrm{un}$) permutations, where, in the latter, at least one quark from the $\rm{t}\bar{\rm{t}}$  decay is not unambiguously matched with a distance of  $\Delta R<0.4$ in the $\eta$-$\phi$ space to any of the four selected jets. In order to check the compatibility of an event with the $\rm{t}\bar{\rm{t}}$ hypothesis and to improve the resolution of the reconstructed quantities, 
the characteristic topology is exploited by the kinematic fit~\cite{Abbott:1998dc}. The fit is repeated for every jet permutation and constrains the mass of two reconstructed W boson to be equal 80.4\,GeV and the mass of $\rm{t}$ quark to be equal to the ones from $\bar{\rm{t}}$. In order to increase the fraction of correct permutations, the goodness-of-fit (gof) probability for the kinematic fit with two degrees of freedom $P_\mathrm{gof} = \exp\left(-\frac{1}{2} \chi^{2}\right)$ is required to be at least 0.2.
Figure~\ref{fig:controlplot-weighted-obs} shows the distributions of the reconstructed mass $m_{\rm{W}}^{\rm{reco}}$ of the W boson decaying to a $\rm{q}\bar{\rm{q}}'$ pair and the top quark mass from the kinematic fit $m_{\rm{t}}^{\rm{fit}}$ for all possible permutations. These two observables are used in the mass extraction.

\vspace{-25pt}

\section{Ideogram Method}
The joint likelihood function is constructed to determine the top quark mass simultaneously with the jet energy scale factor (JSF) in the selected events, where the JSF aims to reduce uncertainty stemming from the standard CMS jet energy corrections~\cite{Khachatryan:2016kdb}.
The observable used for measuring $m_{\rm{t}}$ is the mass $m_{\rm{t}}^{\rm{fit}}$ found in the kinematic fit, while the reconstructed W boson mass $m_{\rm{W}}^{\rm{reco}}$, before it is constrained by the kinematic fit, is used to measure \emph{in situ} the JSF. 
From these distributions, probability density functions $P_{j}$ are derived separately for the different permutation cases $j$: $cp$, $wp$, or $un$. These functions depend on $m_{\rm{t}}$ and JSF and are labeled $P_{j}(m_{\rm{t},i}^{\rm{fit}}|m_{\rm{t}},\mathrm{JSF})$ and  $P_{j}(m_{\rm{W},i}^{\rm{reco}}|m_{\rm{t}},\mathrm{JSF})$ respectively for the $i$th permutation of an event in the final likelihood.
The most likely $m_{\rm{t}}$ and JSF values are obtained by minimizing $-2\ln \left[ \mathcal{L}\left(\rm{sample} | m_{\rm{t}},\mathrm{JSF} \right) \right]$.
The likelihood $\mathcal{L}\left(\rm{sample} | m_{\rm{t}},\mathrm{JSF}\right)$ is defined as:
\begin{equation*}
 \mathcal{L}\left(\rm{sample} | m_{\rm{t}},\mathrm{JSF} \right) = P(\mathrm{JSF}) \cdot  \mathcal{L}\left(m_{\rm{t}},\mathrm{JSF}|\rm{sample}\right) =
 \end{equation*}
\begin{equation*} 
P(\mathrm{JSF}) \cdot \prod_{\rm{events}}\left(\sum_{i=1}^{n} P_{\mathrm{gof}}\left(i\right)\Big(\sum_{j}f_{j}P_{j}(m_{\rm{t},i}^{\rm{fit}}|m_{\rm{t}},\mathrm{JSF})\times P_{j}(m_{\rm{W} ,i}^{\rm{reco}}|m_{\rm{t}},\mathrm{JSF})\Big)\right)^{w_{\rm{evt}}},
\end{equation*}
where $n$ denotes the number of permutations in each event, $j$ labels the permutation cases, and $f_j$ represents their relative fraction. 
The event weight $w_{\rm{evt}}=c\,\sum_{i=1}^{n}P_\mathrm{gof}\left(i\right)$ is introduced to reduce the impact of events without correct permutations, where $c$ normalizes the average $w_{\rm{evt}}$ to 1.
The choice of the prior $P(\mathrm{JSF})$ in the likelihood fit defines the three set of the solution: 1D approach that estimates only the most probable $m_{\rm{t}}$, since the JSF is fixed to unity, 2D approach with an unconstrained JSF estimates both $m_{\rm{t}}$ and JSF. Finally, in the "`hybrid"' approach uses the prior knowledge on the JSF derived from the uncertainties on the jet energy corrections~\cite{Khachatryan:2016kdb} and computes the prior $P(\mathrm{JSF})$ of a Gaussian form centered at 1.0 with a width $\sigma_{\mathrm{prior}}$ calculated from the relative weight $w_\mathrm{hyb} = 0.3$. This results in a width for $P(\mathrm{JSF})$ of ${\sigma_{\mathrm{prior}} = \delta\rm{JSF}^{\rm{2D}}_{\text{stat}} \sqrt{1/w_\mathrm{hyb} - 1} = 0.0012}$, where $\delta\rm{JSF}^{\rm{2D}}_{\rm{stat}}$ is the statistical uncertainty on the JSF obtained from the 2D method and is expected to be $0.0008$. 

Due to some simplifications in the used ideogram method, like fixed permutation number and neglecting of the background, the 2D method is calibrated. This is performed by conducting  10\,000 pseudo-experiments for each combination of the seven $m_{\rm{t},\text{gen}}$  and the five $\mathrm{JSF}$ values for the muon and electron channels separately using simulated $\rm{t}\bar{\rm{t}}$ and background events. Corrections for the statistical uncertainty of the method are derived from the widths of the corresponding pull distributions and have a size of 5\%. 
\vspace{-15pt}
\section{Systematic uncertainties}
The results on the systematic uncertainty estimation is shown in Table~\ref{tab:Systematic-uncertainties}. Only the largest systematic uncertainties are discussed in this work, more detailed information is given in~\cite{CMS-PAS-TOP-17-007}.
\begin{table}[!htb]
{\footnotesize
\vspace{-15pt}
\caption{\label{tab:Systematic-uncertainties}List of systematic uncertainties for the fit to the combined data set.} 
\begin{center}
\begin{tabular}{l|cc|c|cc}
 & \multicolumn{2}{c|}{2D approach} & 1D approach & \multicolumn{2}{c}{Hybrid} \\
\hline 
  & (GeV) & & (GeV) & (GeV)& \\
\hline 
\hline 
Experimental uncertainties &  &  &  &  & \\
\hline 
JEC (quad. sum) & 0.13 & 0.002 & 0.85 & 0.19 & 0.003 \tabularnewline 
JER  & 0.08 & 0.001 & 0.04 & 0.04 & 0.001\\
Pileup  & 0.08 & 0.001 & 0.02 & 0.05 & 0.001\\
b tag. \& non-t$\bar{\rm{t}}$ BG  & 0.07 & 0.001 & 0.05 & 0.06& 0.001 \\
\hline 
\hline 
Modeling uncertainties &  &  &  & &\\
\hline 
JEC: Flavor (linear sum) & 0.42 & 0.001 & 0.31 & 0.39&$<$0.001 \tabularnewline
b-jet modeling (quad. sum)  & 0.13 & 0.001 & 0.09 & 0.12&$<$0.001 \tabularnewline 
\hline 
ME generator  & 0.19  & 0.001  & 0.29  & 0.22 & 0.001 \\ 
ISR \& FSR PS scale & 0.25 & 0.004 & 0.24 & 0.14 & 0.003  \\
PDF, $\mu_{\rm{R}}$ \& $\mu_{\rm{F}}$, ME/PS & 0.08 & 0.001 & 0.04 & 0.05 & 0.001\\
\hline 
Underlying event  & 0.07 & 0.001 & 0.10 & 0.06 &  0.001 \\
Early resonance decays & 0.22 & 0.008 & 0.23 & 0.03 & 0.005 \\
CR modeling & 0.34 & 0.001 & 0.42 & 0.31 & 0.001 \\
\hline 
\hline 
\textbf{Total systematic} &  \textbf{0.71} & \textbf{0.010} & \textbf{1.09} & \textbf{0.62 }& \textbf{0.008}\\
\hline
Statistical (expected) & 0.09 & 0.001 & 0.05 & 0.07 & 0.001  \\ 
\hline 
\hline 
\textbf{Total (expected)} & \textbf{0.72} & \textbf{0.010} & \textbf{1.09} & \textbf{0.62}  & \textbf{0.008}\\
\end{tabular}
\end{center}

}
\end{table}

The systematic uncertainties are larger than for the Run~1 result of $m_{\rm{t}} = 172.35\pm 0.16\,\rm{(stat+JSF)} \pm 0.48\,\rm{(syst)}$\,GeV, due to the more evolved treatment of the modeling uncertainties. This is mainly caused by the evaluation of a broader set of color reconnection models that were not available in Run~1. In Run\,2 the effects of color reconnection on the top quark decay products can be turned on in $\textsc{pythia}$\,8 by enabling early resonance decays (ERD). In the default sample the early resonance decays are turned off. Additionally, the uncertainties that arise from ambiguities in modeling color reconnection effects are estimated by comparing the default model in $\textsc{pythia}$\,8 with 
two alternative models of color reconnection, a model with string formation beyond leading color (``QCD inspired'')~\cite{Christiansen:2015yqa} and a model in that the gluons can be moved to another string (``gluon move'')~\cite{Argyropoulos:2014zoa}. The influence of the matrix-element generator is estimated by using a sample from the a$\textsc{mc@nlo}$ generator and the \textsc{FxFx} matching~\cite{Frederix:2012ps} instead of the $\textsc{powheg}$~v2 generator used as default. The systematic uncertainty coming from the jet energy correction is still the dominant one. 

\vspace{-15pt}
\section{Results}
Out of $35.9~\mathrm{fb}^{-1}$ of 2016 data, 161\,496 lepton+jets events are selected, which yields for the 2D method:
\begin{eqnarray*}
m_{\rm{t}}^{\text{2D}} & = & 172.40\pm0.09\mbox{ (stat+JSF)}\pm0.68\mbox{ (syst) GeV},\\
\mbox{JSF}^{\text{2D}} & = & 0.994\pm0.001\mbox{ (stat)}\pm0.010\mbox{ (syst)},
\end{eqnarray*}
for the 1D method:
\begin{eqnarray*}
m_{\rm{t}}^{\text{1D}} & = & 171.93\pm0.07\mbox{ (stat)}\pm1.09\mbox{ (syst) GeV},
\end{eqnarray*}
and for the hybrid method:
\begin{eqnarray*}
m_{\rm{t}}^{\text{hyb}} & = & 172.25\pm0.08\mbox{ (stat+JSF)}\pm0.62\mbox{ (syst) GeV},\\
\mbox{JSF}^{\text{hyb}} & = & 0.996\pm0.001\mbox{ (stat)}\pm0.008\mbox{ (syst)}.
\end{eqnarray*}
As expected, the hybrid method offers the lowest overall uncertainty. 
Due to the larger integrated luminosity and the higher $\rm{t}\bar{\rm{t}}$ cross section at $\sqrt{s}=13~\mathrm{TeV}$ the statistical uncertainty is halved compared to the Run~1 result.
\vspace{-15pt}
\section{Measured top quark mass as a function of kinematic observables}
\begin{figure*}[!htb]
\centering
  \includegraphics[width=0.32\textwidth]{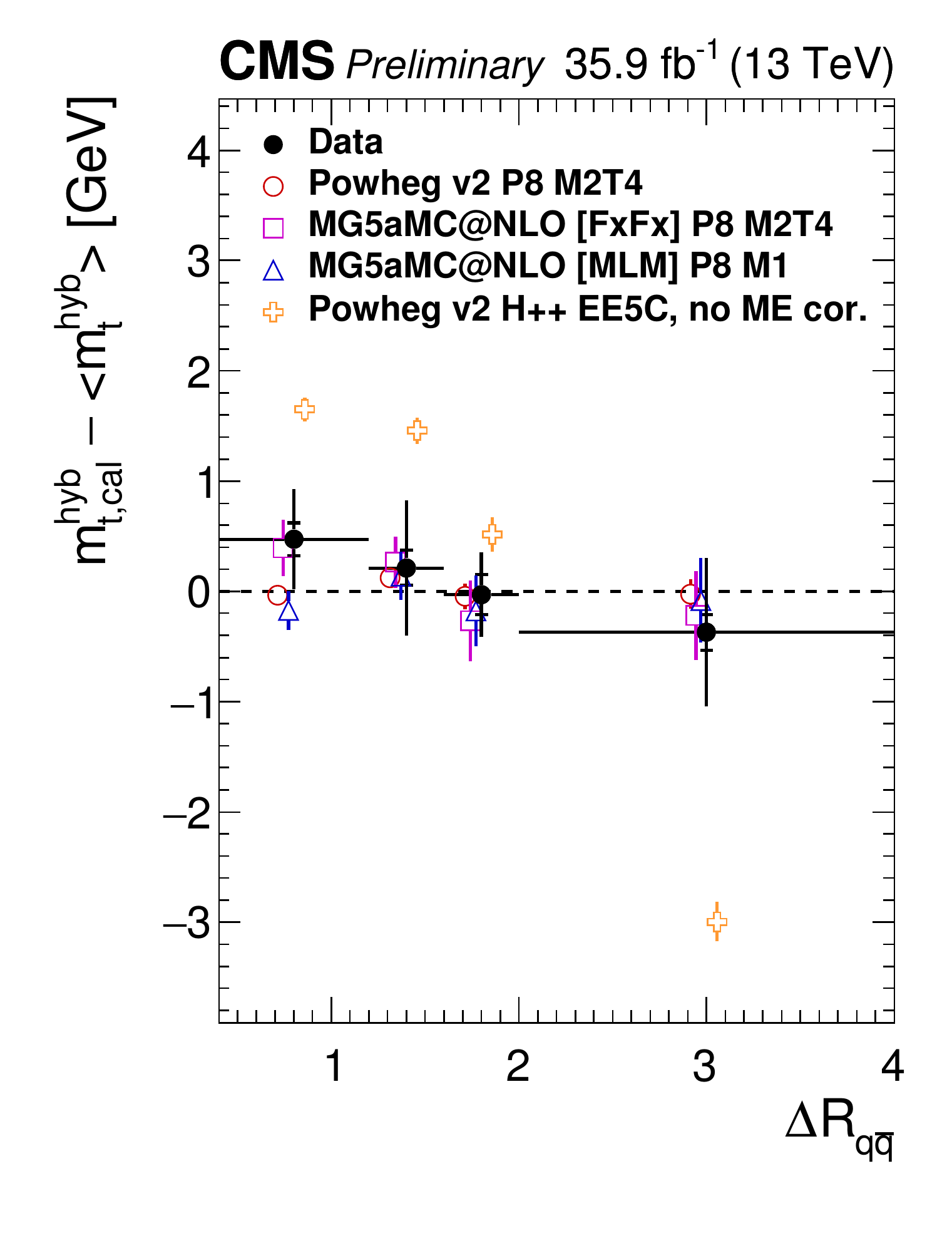}
  \includegraphics[width=0.32\textwidth]{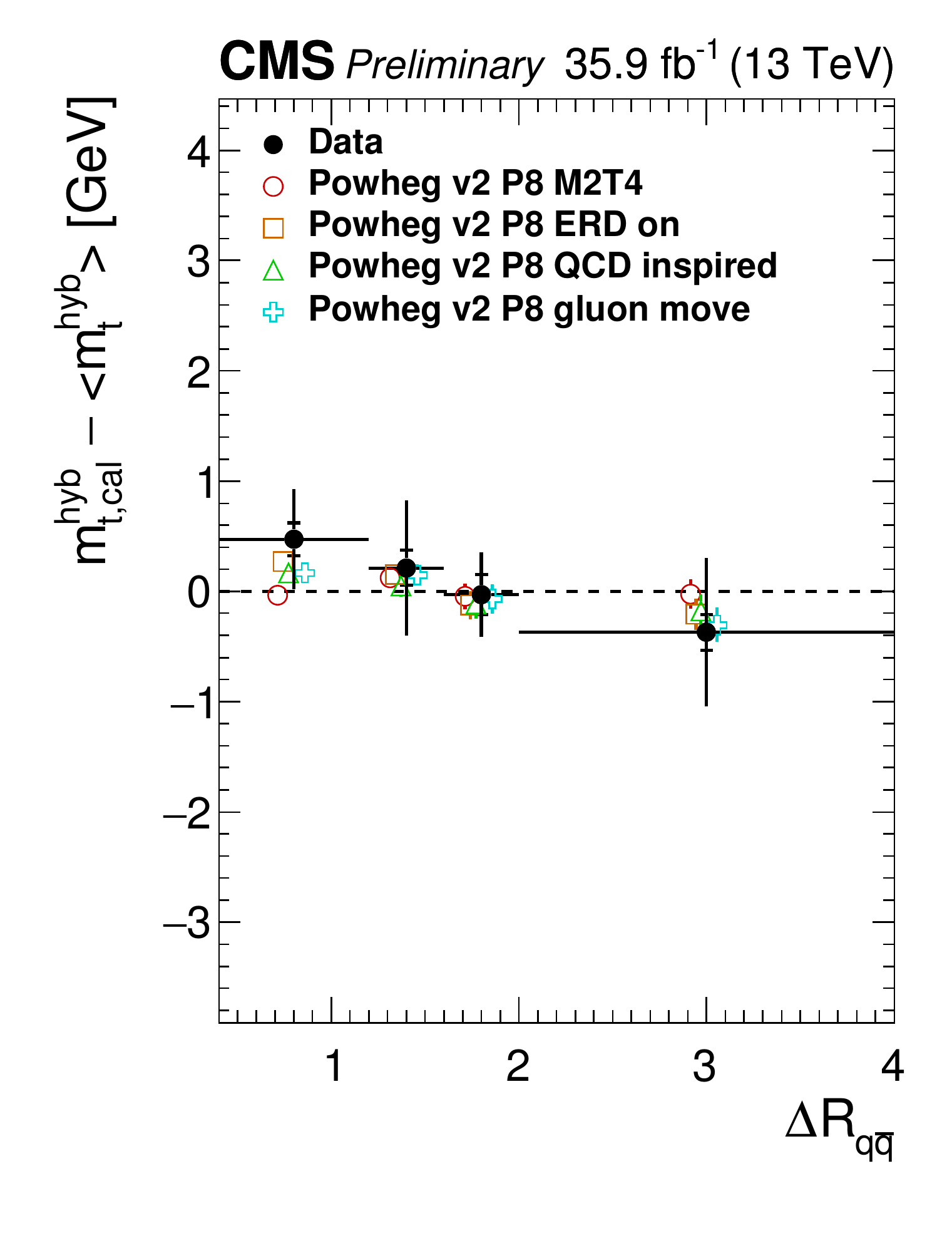}
  \includegraphics[width=0.32\textwidth]{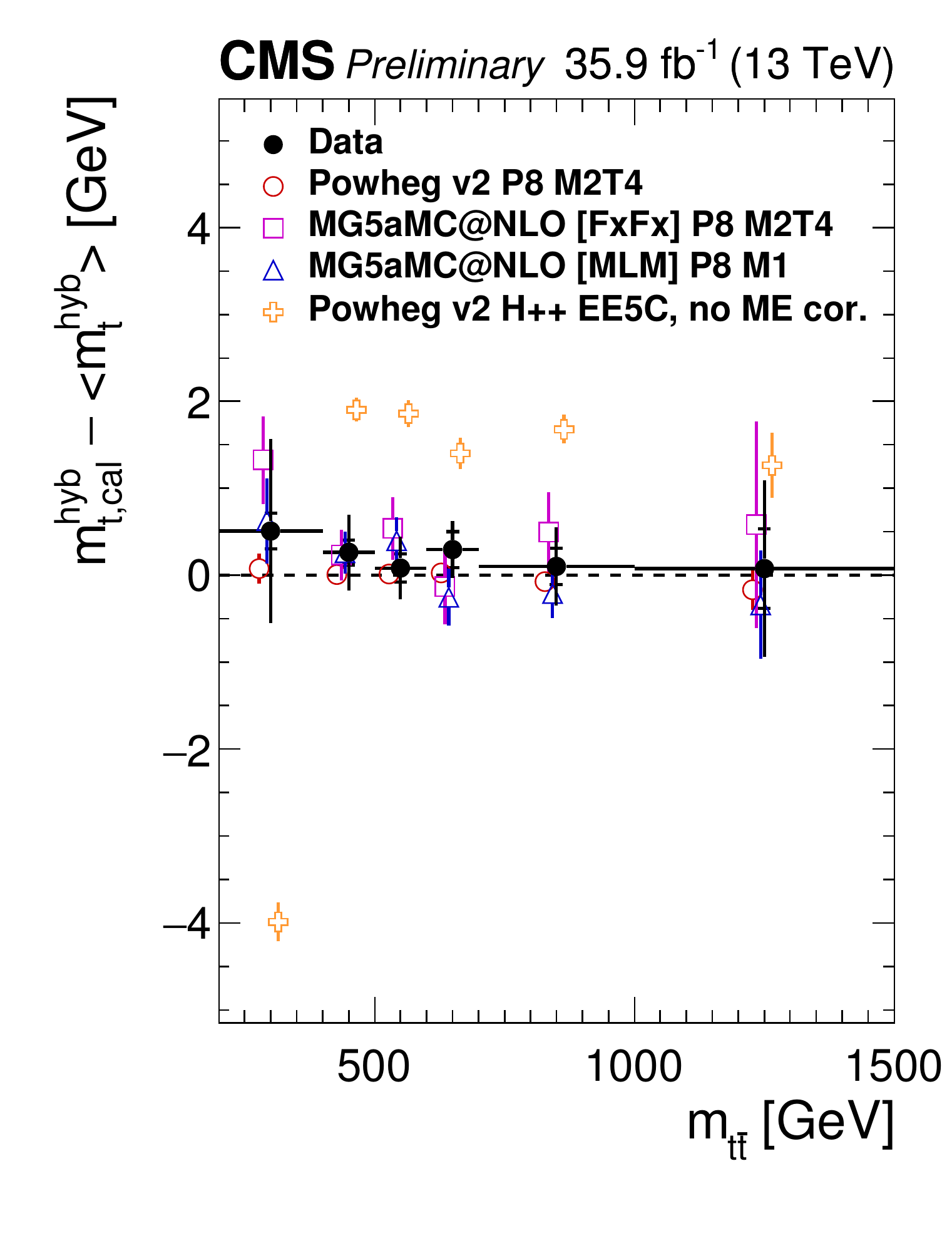}
  \caption{Measurements of $m_{\rm{t}}$ as a function of the $\Delta R$ between the light quark jets ($\Delta R_{\rm{q}\bar{\rm{q}}}$) compared to different generator (left) and different CR models (middle) choices, and $m_{\rm{t}}$ as a function of the invariant mass of the $\rm{t}\bar{\rm{t}}$ system ($m_{\rm{t}\bar{\rm{t}} }$) compared to different generator (right). Figure taken from~\cite{CMS-PAS-TOP-17-007}.}
  \label{fig:diff_top_gen}
\end{figure*}

The comparison of different models for soft and perturbative QCD are the main source of systematic uncertainties on the presented analysis. Differential measurements of $m_{\rm{t}}$ as a function of the kinematic properties of the $\rm{t}\bar{\rm{t}}$ system allow the validation of the different models and identification of possible biases in the presented measurement.
Variables are selected that probe potential effects from color reconnection, initial- and final-state radiation, and the kinematics of the jets coming from the top quark decays.
Figure~\ref{fig:diff_top_gen} demonstrates the dependence of the measured $m_{\rm{t}}$ on the  $\Delta R$ between the light quark jets ($\Delta R_{\rm{q}\bar{\rm{q}}}$) and on  the invariant mass of the $\rm{t}\bar{\rm{t}}$ system ($m_{\rm{t}\bar{\rm{t}} }$) compared to different generator and different CR models choices. Even with the improved statistical precision from the new data, no significant deviation in the value of the measured $m_{\rm{t}}$ is observed. The results for $\textsc{powheg}$~v2 + \textsc{herwig}$++$~\cite{Bahr:2008pv} differ from the data and all other setups, which might be explained by the missing matrix-element corrections to the top quark decay.

\vspace{-25pt}
\section{Summary}
This result of $m_{\rm{t}} = 172.25\pm 0.08\,\rm{(stat+JSF)} \pm 0.62\,\rm{(syst)}$\,GeV is the first result of the top quark mass measured with full 2016 data and the new NLO generator setups. The evaluation of a broader set of color reconnection models for the systematic uncertainty estimation is done. This measurement is consistent with the Run~1 result. No shift in the measured top quark mass from the new simulation at next-to-leading order with $\textsc{powheg}$~v2 and $\textsc{pythia}$\,8 and the new experimental setup is observed.
 \bibliographystyle{unsrt}
\bibliography{bibliography}









\end{document}